%% file: Template_ISBI_latex.tex
\documentclass{article}
\usepackage{spconf,amsmath,graphicx}
\usepackage{color} 

\usepackage{enumitem}
\setlist{nosep, leftmargin=14pt}

\title{Multi-center anatomical segmentation with heterogeneous labels via landmark-based models}

\name{Nicolás Gaggion$^{\star}$ \qquad Maria Vakalopoulou$^{\dagger}$ \qquad Diego H. Milone$^{\star}$ \qquad Enzo Ferrante$^{\star}$} 
\address{$^{\star}$ Research Institute for Signals, Systems and Comp. Intelligence, sinc(i), CONICET-UNL, Argentina \\
$^{\dagger}$ MICS, CentraleSupélec, Université Paris-Saclay, Inria Saclay, France}

\usepackage{graphicx}
\usepackage{multirow}
\usepackage[table,xcdraw]{xcolor}
\usepackage{hyperref}       
\usepackage{url}            

\begin{document}
\maketitle
\begin{abstract}
  Learning anatomical segmentation from heterogeneous labels in multi-center datasets is a common situation encountered in clinical scenarios, where certain anatomical structures are only annotated in images coming from particular medical centers, but not in the full database. Here we first show how state-of-the-art pixel-level segmentation models fail in naively learning this task due to domain memorization issues and conflicting labels. We then propose to adopt HybridGNet, a landmark-based segmentation model which learns the available anatomical structures using graph-based representations. By analyzing the latent space learned by both models, we show that HybridGNet naturally learns more domain-invariant feature representations, and provide empirical evidence in the context of chest X-ray multiclass segmentation. We hope these insights will shed light on the training of deep learning models with heterogeneous labels from public and multi-center datasets.
\end{abstract}


\section{Introduction}

Anatomical segmentation is one of the pillar problems in medical image analysis, required by several downstream tasks like radiotherapy planning \cite{ferrante2018weakly} or shape variability analysis in computational anatomy \cite{cerrolaza2019computational}. 
Fully convolutional neural networks such as UNet \cite{ronneberger2015u}, and its self-configuring variant nnUNet \cite{isensee2021nnunet}, have become the state-of-the-art for this task. In this work, we are interested in addressing two common situations encountered when training anatomical segmentation models in real clinical scenarios: multi-center image databases and heterogeneous labels. On one hand, multi-center databases may lead to domain shift problems \cite{guan2021domain} due to changes in intensity distribution caused by differences in acquisition device or protocol parameters. On the other hand, heterogeneous or missing labels \cite{kemnitz2018combining} make it difficult to train a single segmentation model for all regions of interest (ROIs), as missing labels in different images may send contradictory training signals. Notably, when we face both issues at the same time, the problem is far from trivial as it lies in the intersection of multi-task learning, domain adaptation and weakly supervised learning \cite{dorent2021learning}. As we will show in this work, when different organs are annotated in images coming from various centers, commonly used pixel-level segmentation methods like UNet and nnUNet trained with standard procedures tend to associate certain labels to specific domains. 

Several methods have been proposed to independently address the problems of domain shift \cite{guan2021domain,palladino2020unsupervised,marin2022} and heterogeneous labels \cite{petit2018handling,kemnitz2018combining,filbrandt2021learning} in medical image segmentation. As for the joint problem, Dorent and coworkers \cite{dorent2021learning} proposed a framework which combines a variational formulation to cope with heterogeneous labels, with conventional techniques based on data augmentation, adversarial learning, and pseudo-healthy image generation to address domain shift. 
In this work, we argue that landmark based segmentation methods like the HybridGNet \cite{gaggion2021,gaggion2022} can naturally handle these scenarios, as they incorporate prior knowledge about the expected anatomy, without additional burden related to data augmentation, adversarial training, or image generation. We first provide empirical evidence showing how widely used pixel-level approaches drastically fail to learn robust segmentation models using heterogeneous labels from multicentric datasets, while HybridGNet can naturally handle this problem, avoiding memorization issues. Further analysis of the latent spaces learned by the different architectures, indicates that generative landmark-based approaches like HybridGNet tend to learn more invariant representations, which helps to improve the robustness with respect to domain memorization.

\begin{figure*}[t!]
    \centering
    \includegraphics[width=\linewidth]{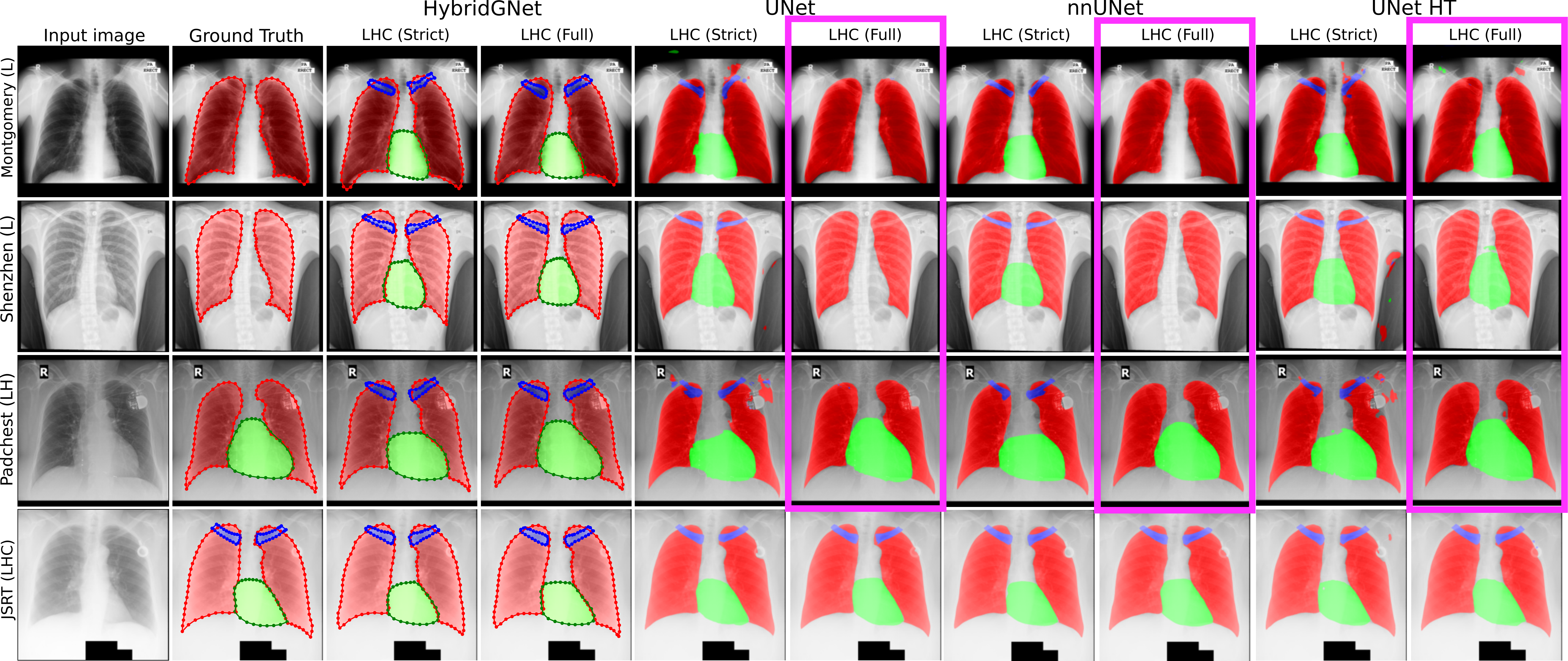}
    \caption{\textbf{Qualitative examples}. Examples from 4 datasets (rows) segmented with different methods and annotation settings (columns). Note how pixel UNet and nnUNet (in pink) fail completely to segment structures that are not presented in the corresponding dataset when trained with heterogeneous labels (i.e. LHC (Full)). It is not the case for the HybridGNet, which always provides segmentations of all structures. Note also that the UNet HT (trained using the same heterogeneous setup as HybridGNet) produces heart segmentation for all datasets, as there are no conflicting labels. However, it only segments clavicles for JSRT, due to the conflicting annotations in the overlapping area between lung and clavicles (see Section 3 for more details).}
    \label{fig:quali}
\end{figure*}

\section{Methods and experiments}

\noindent \textbf{Problem statement and experimental setup:} 
We explore anatomical segmentation of lung, heart and clavicles in a multi-center database of chest X-ray images, with heterogeneous labels. The database is composed of 4 publicly available datasets (JSRT \cite{jsrt_shiraishi2000development}, Padchest \cite{bustos2020padchest}, Montgomery \cite{montgomeryset} and Shenzhen \cite{shenzhen}), which originally provide pixel level annotations. Since the proposed method employs landmark-based annotations, we adopted the publicly available \textit{Chest X-ray landmark dataset}, which provides landmarks for 3 different organs from the aforementioned databases (\href{https://github.com/ngaggion/Chest-xray-landmark-dataset}{github.com/ngaggion/Chest-xray-landmark-dataset}). Images from JSRT (246 subjects) include annotations for lungs, heart and clavicle (LHC); Padchest (137 subjects) include lungs and heart (LH); while Montgomery (138 subjects) and Shenzhen (390 subjects) include only lung (L). To evaluate each domain separately, we divide the datasets into 80\% train/val and 20\% test partitions.

\noindent \textbf{HybridGNet:} HybridGNet is a landmark-based segmentation model, where the ROIs are encoded as anatomical graphs representing the organ contour. It follows an encoder-decoder architecture that combines standard convolutions for image encodings, with graph generative models to extract anatomically plausible representations directly from images. The model is trained to minimize the mean squared error (MSE) between the predicted node positions and the ground truth coordinates. Pixel-level masks are then generated by filling in the contours. See \cite{gaggion2021,gaggion2022} for more details about HybridGNet.

\noindent \textbf{Training landmark-based models with heterogeneous labels:}
HybridGNet provides a natural way to learn with heterogeneous labels, which only relies on indexation. The model outputs a $D \times 2$ matrix, where the number of nodes $D$ is fixed and sequentially ordered: first lung nodes, $t_{L}$, then heart nodes, $t_{H}$, and finally clavicles nodes, $t_{C}$, as in the ground-truth $target = [t_{L}, t_{H}, t_{C}]$. 
The length of every subset is $D_{L}$, $D_{H}$, $D_{C}$, respectively. 
For training, batches composed of images \textit{from a single database at a time} are randomly chosen at every iteration, thus constraining the type of annotation to those available for that dataset. For example, if the input batch only includes lungs, the loss function for that gradient descent iteration is only evaluated for the first $D_{L}$ nodes, and errors are not back-propagated for heart and clavicle. The same is done for the LH task, where the loss is evaluated in the first $D_{L}+D_{H}$ nodes, ignoring the rest of the output.
This is implemented via slicing operations, and constitutes the only modification made to HybridGNet.
\input{table1}

\noindent \textbf{Baselines and heterogeneous UNet training:}
We propose to compare the HybridGNet with two pixel-level segmentation models: a UNet \cite{ronneberger2015u} with residual convolutional blocks, trained with a compound cross entropy and soft Dice loss \cite{milletari2016v}; and a nnUNet \cite{isensee2021nnunet} trained with its self-configuring method. We also propose to train the UNet in the same heterogeneous training (HT) setup as the HybridGNet for fair comparison. In UNet HT, each training batch contains a specific set of labels, and we avoid back-propagating the gradient loss for unseen labels in the batch. We implement this method simply treating each anatomical structure as an independent binary segmentation problem. The UNet HT model thus has one independent output feature map per anatomical structure, akin to a multi-label classification problem. We apply a sigmoid non-linearity to each output segmentation map. We use binary cross-entropy and a modified soft Dice loss function to allow using a single feature map, instead of the standard one-hot encoding used when classes are mutually exclusive. At test time, the sigmoid outputs are just thresholded at 0.5, obtaining an independent binary map for each structure.

\noindent \textbf{Training details:} 
As not all labels are available for each dataset, we took into account the label availability and devised two different training settings: \textbf{Strict} and \textbf{Full}. While \textbf{Strict} indicates that only datasets annotated with the particular listed labels were used, \textbf{Full} indicates that all datasets were used for training (resulting in an heterogeneous label setting). Thus, we trained models in the following settings:

\begin{itemize}

\item \textbf{Strict training setting}: 

    \begin{itemize}
        \item \textbf{LH}: Models were trained to predict lung and heart, using only images that had both lung and heart annotations (i.e. JSRT and Padchest datasets).
        \item \textbf{LHC}: Models were trained to predict lung, heart and clavicles, using only images that had all annotations available (thus, just JSRT dataset). 
    \end{itemize}

\item \textbf{Full training setting:}
\begin{itemize}
        
        \item \textbf{L}: Models were trained with all datasets to predict only lungs. This case can also be considered \textbf{Strict}, since all datasets have lung annotations available.
        \item \textbf{LH}: Models were trained to predict lungs and heart using all datasets in an heterogeneous annotation setting, i.e. some images had only lung annotations (Montgomery and Shenzhen) and others had lung and heart (JSRT and Padchest).
        \item \textbf{LHC}: Models were trained to predict lungs, heart and clavicles using all datasets in an heterogeneous setting, i.e. some images had only lung annotations (Montgomery and Shenzhen), some had lung and heart (Padchest), while some had the 3 structures (JSRT).
    \end{itemize}
\end{itemize}

\noindent \textbf{Artificial removal of labels:}
As we will discuss in the next section, our experiments show that naively trained pixel-level approaches learn to map anatomical structures to datasets where they were annotated, failing to segment them in datasets where these structures are not labeled. However, as we do not have ground-truth for these structures, we cannot show this quantiatively. To overcome this limitation and provide quantitative support for our claims, we took the two datasets with more than one annotated structure (JSRT and Padchest), and artificially removed one of the organs during training.  This resulted in 4 different experiments, where we can compute quantiative results for labels that were not seen during training: removing lungs from JSRT (\textbf{Exp 1}), removing heart from JSRT (\textbf{Exp 2}), removing lungs from Padchest (\textbf{Exp 3}) and removing heart from Padchest (\textbf{Exp 4}).

\section{Results, discussion and conclusions} 

Figure \ref{fig:quali} shows one of our main findings: when trained with heterogeneous labels associated to different datasets (i.e. LH (Full) and LHC (Full)), naive pixel-based methods segment only those ROIs that were annotated in the corresponding dataset (see first two cases highlighted in pink). Meanwhile, the UNet HT model which is aware of the heterogeneous labels by ignoring annotations not present in every specific dataset, only segments classes that are not in conflict 
(i.e. we say a pixel class is in conflict if it is considered to be part of one class for a specific dataset, but also part of a different class in other datasets, like clavicles). Instead, the HybridGNet always predicts the complete set of anatomically plausible segmentations. This effect is present on all the samples for each dataset. UNet and nnUNet clearly memorize which structure was annotated in what dataset, and use the multi-centric distribution shift as a shortcut to decide what ROIs should be segmented in every test dataset. UNet HT overcomes this issue for the heart masks, as we make it aware of heterogeneous labels by ignoring classes that are not annotated in specific datasets, but still fails with the clavicles, as they conflict with lung labels. On the contrary, HybridGNet is forced to predict all ROIs by construction, using the anatomical priors encoded in the learned latent space to infer the organ position, even if it was not present in that particular dataset.

\begin{figure}[t]
    \centering
    \includegraphics[width=\linewidth]{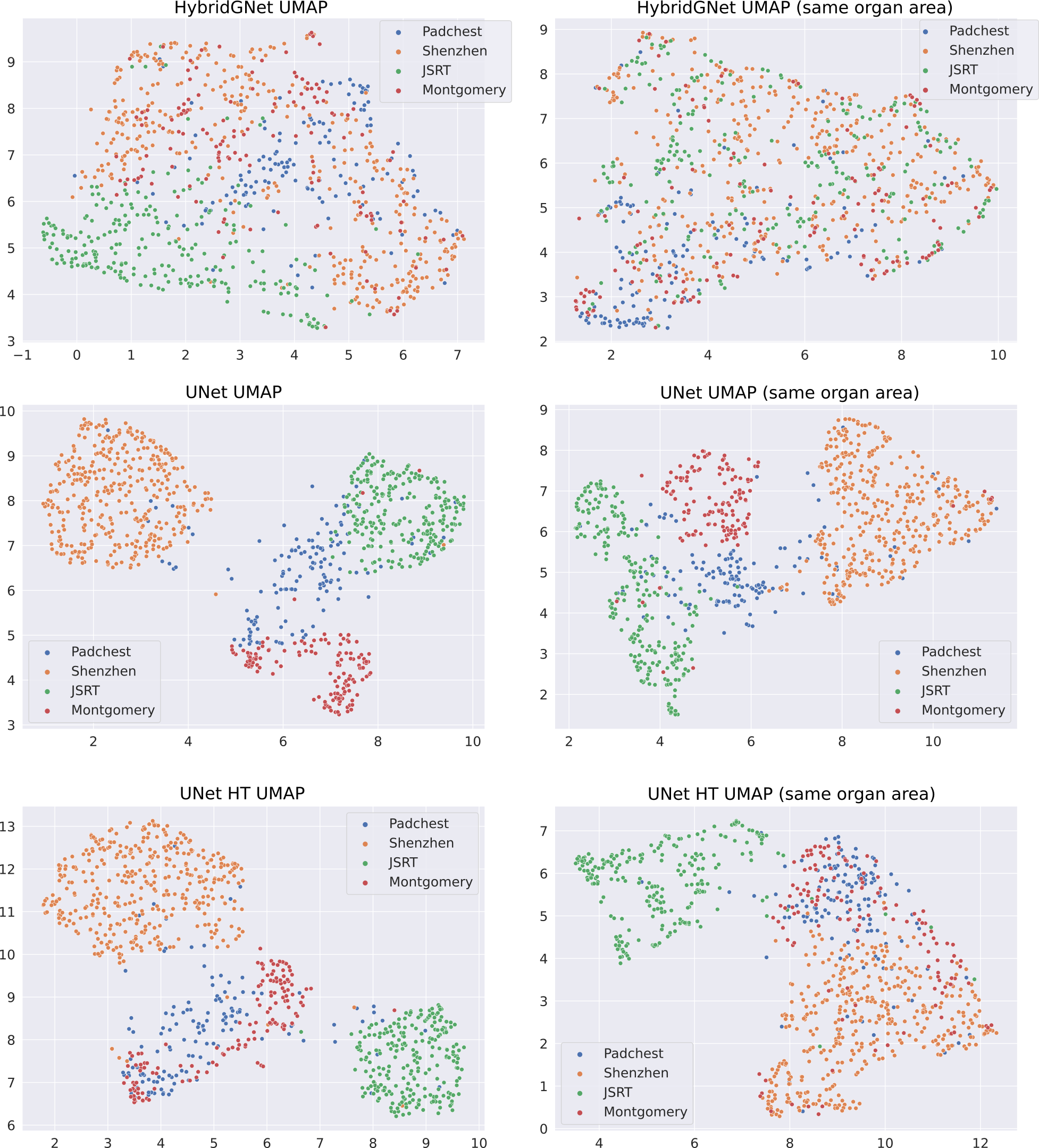}
    \caption{\textbf{Latent space inspection via UMAP embeddings:}
    Different datasets are shown in colors, allowing to see how both UNet and UNet HT models tend to clusterize images per dataset, while HybridGNet doesn't, explaining the improved robustness to domain-label memorization.}
    \label{fig:umap}
\end{figure}
\input{table2}

To better understand the reasons behind domain memorization, we performed dimensionality reduction on both the latent space of the HybridGNet model and the bottleneck features of the UNets. We analyzed the \textbf{LHC (Full)} models, and used  UMAP\cite{mcinnes2018umap} for dimensionality reduction. Figure \ref{fig:umap} (Left column) shows the 2D projection of images from all datasets for each model. UNet and UNet HT clearly clusterize samples per dataset. Since JSRT images tend to have much bigger lung area than the other datasets, we also experimented scaling all images to have the same organ's bounding box area, discarding the potential clustering effect associated to the lung size and not to the multi-center intensity shift. This is shown in Figure \ref{fig:umap} (Right column), where the clustering was completely removed on the HybridGNet latent space, but it did not affect the UNet and UNet HT. nnUNet results could not be obtained due to the encapsulation of the training and test framework, but are expected to behave similarly to the naive UNet since both models share the same underlying architecture. This clustering effect explains why memorization issues happen in UNet and UNet HT, but not in HybridGNet.

Table \ref{table:table_nnunet} shows quantitative results for the different training scenarios, models and test sets. In general, we see that incorporating the Full dataset through heterogeneous labels improve performance when compared to the Strict case, which only employs images where all annotations are available. Performance is only evaluated for ROIs available as ground-truth in every dataset. As shown in Figure \ref{fig:quali}, for the heterogenous settings (\textbf{LHC (Full)} and \textbf{LH (Full)}) UNet and nnUNet drastically fail at segmenting structures that were not present during training in the corresponding dataset. This is due to the memorization issues and conflicting background/organ labels in different domains. This would imply a Dice of 0 for UNet and nnUNet if ground-truth were available for evaluation.

To quantify these results, we artificially removed labels from datasets with more than one annotated structure (JSRT and Padchest). Table \ref{table:table_experiment} shows quantitative results for each model, on the same test sets that Table \ref{table:table_nnunet}. While naive segmentation models (UNet and nnUNet) drastically fail to segment missing structures in the corresponding datasets (showing Dice of around $0$), the methods that are aware of heterogeneous labels (HybridGNet and UNet HT) slightly reduce their performance, but can still recover the anatomical structure. More importantly, note that HybridGNet largely outperforms UNet HT in terms of HD distance for the missing labels (in red), while it is competitive in terms of Dice coefficient.

\noindent \textbf{Conclusions.} Here we show how HybridGNet can deal with heterogeneous labels in multi-center scenarios, where state-of-the-art UNet and nnUNet drastically fail. Moreover, the UNet HT model trained to be aware of heterogeneous labels by ignoring missing structures proved to be useful in avoiding contradictory signals between missing annotations and the background. However, this is not enough when there are contradictory signals between organs. In these cases, landmark-based models like HybridGNet offer a simple framework which can easily handle overlapping structures, particularly heterogeneous labels in multi-center scenarios.

\section{Compliance with ethical standards}


This is a numerical simulation study for which no ethical approval was required. 

\section{Acknowledgements}

The authors gratefully acknowledge NVIDIA Corporation with the donation of the GPUs used for this research, and the support of UNL (CAID-PIC50420150100098LI, CAID-PIC-50220140100084LI) and ANPCyT (PICT 2016-0651, PICT 2018-03907).

\bibliographystyle{IEEEbib}
\bibliography{bib.bib}

\end{document}

%% file: table1.tex

\begin{table*}[t!]
\resizebox{\textwidth}{!}{\begin{tabular}{cc|ccc|ccc|cccccc|ccccccccc}
\hline
 &
   &
  \multicolumn{3}{c|}{\textbf{Montgomery}} &
  \multicolumn{3}{c|}{\textbf{Shenzhen}} &
  \multicolumn{6}{c|}{\textbf{Padchest}} &
  \multicolumn{9}{c}{\textbf{JSRT}} \\
 &
   &
  \multicolumn{3}{c|}{\textbf{Lungs}} &
  \multicolumn{3}{c|}{\textbf{Lungs}} &
  \multicolumn{3}{c}{\textbf{Lungs}} &
  \multicolumn{3}{c|}{\textbf{Heart}} &
  \multicolumn{3}{c}{\textbf{Lungs}} &
  \multicolumn{3}{c}{\textbf{Heart}} &
  \multicolumn{3}{c}{\textbf{Clavicles}} \\
\multirow{-3}{*}{\textbf{Model}} &
  \multirow{-3}{*}{\textbf{Trained in}} &
  \textbf{MSE} &
  \textbf{Dice} &
  \textbf{HD} &
  \textbf{MSE} &
  \textbf{Dice} &
  \textbf{HD} &
  \textbf{MSE} &
  \textbf{Dice} &
  \textbf{HD} &
  \textbf{MSE} &
  \textbf{Dice} &
  \textbf{HD} &
  \textbf{MSE} &
  \textbf{Dice} &
  \textbf{HD} &
  \textbf{MSE} &
  \textbf{Dice} &
  \textbf{HD} &
  \textbf{MSE} &
  \textbf{Dice} &
  \textbf{HD} \\ \hline
 &
  \cellcolor[HTML]{C9DAF8}L (Both) &
  \cellcolor[HTML]{C9DAF8}128.4 &
  \cellcolor[HTML]{C9DAF8}0.97 &
  \cellcolor[HTML]{C9DAF8}27.4 &
  \cellcolor[HTML]{C9DAF8}142.9 &
  \cellcolor[HTML]{C9DAF8}0.97 &
  \cellcolor[HTML]{C9DAF8}32.5 &
  \cellcolor[HTML]{C9DAF8}152.1 &
  \cellcolor[HTML]{C9DAF8}0.96 &
  \cellcolor[HTML]{C9DAF8}36.6 &
  \cellcolor[HTML]{C9DAF8}- &
  \cellcolor[HTML]{C9DAF8}- &
  \cellcolor[HTML]{C9DAF8}- &
  \cellcolor[HTML]{C9DAF8}120.2 &
  \cellcolor[HTML]{C9DAF8}0.97 &
  \cellcolor[HTML]{C9DAF8}33.0 &
  \cellcolor[HTML]{C9DAF8}- &
  \cellcolor[HTML]{C9DAF8}- &
  \cellcolor[HTML]{C9DAF8}- &
  \cellcolor[HTML]{C9DAF8}- &
  \cellcolor[HTML]{C9DAF8}- &
  \cellcolor[HTML]{C9DAF8}- \\
 &
  \cellcolor[HTML]{F4CCCC}LH (Strict) &
  \cellcolor[HTML]{F4CCCC}295.6 &
  \cellcolor[HTML]{F4CCCC}0.95 &
  \cellcolor[HTML]{F4CCCC}38.5 &
  \cellcolor[HTML]{F4CCCC}264.4 &
  \cellcolor[HTML]{F4CCCC}0.95 &
  \cellcolor[HTML]{F4CCCC}43.1 &
  \cellcolor[HTML]{C9DAF8}201.8 &
  \cellcolor[HTML]{C9DAF8}0.95 &
  \cellcolor[HTML]{C9DAF8}41.0 &
  \cellcolor[HTML]{C9DAF8}351.7 &
  \cellcolor[HTML]{C9DAF8}0.94 &
  \cellcolor[HTML]{C9DAF8}36.8 &
  \cellcolor[HTML]{C9DAF8}155.7 &
  \cellcolor[HTML]{C9DAF8}0.97 &
  \cellcolor[HTML]{C9DAF8}36.3 &
  \cellcolor[HTML]{C9DAF8}382.4 &
  \cellcolor[HTML]{C9DAF8}0.94 &
  \cellcolor[HTML]{C9DAF8}34.5 &
  \cellcolor[HTML]{C9DAF8}- &
  \cellcolor[HTML]{C9DAF8}- &
  \cellcolor[HTML]{C9DAF8}- \\
 &
  \cellcolor[HTML]{D9EAD3}LH (Full) &
  \cellcolor[HTML]{D9EAD3}104.4 &
  \cellcolor[HTML]{D9EAD3}0.97 &
  \cellcolor[HTML]{D9EAD3}27.8 &
  \cellcolor[HTML]{D9EAD3}145.9 &
  \cellcolor[HTML]{D9EAD3}0.97 &
  \cellcolor[HTML]{D9EAD3}32.8 &
  \cellcolor[HTML]{D9EAD3}187.0 &
  \cellcolor[HTML]{D9EAD3}0.96 &
  \cellcolor[HTML]{D9EAD3}38.8 &
  \cellcolor[HTML]{D9EAD3}342.0 &
  \cellcolor[HTML]{D9EAD3}0.94 &
  \cellcolor[HTML]{D9EAD3}36.0 &
  \cellcolor[HTML]{D9EAD3}124.8 &
  \cellcolor[HTML]{D9EAD3}0.97 &
  \cellcolor[HTML]{D9EAD3}31.0 &
  \cellcolor[HTML]{D9EAD3}375.8 &
  \cellcolor[HTML]{D9EAD3}0.94 &
  \cellcolor[HTML]{D9EAD3}33.8 &
  \cellcolor[HTML]{D9EAD3}- &
  \cellcolor[HTML]{D9EAD3}- &
  \cellcolor[HTML]{D9EAD3}- \\
 &
  \cellcolor[HTML]{F4CCCC}LHC (Strict) &
  \cellcolor[HTML]{F4CCCC}571.5 &
  \cellcolor[HTML]{F4CCCC}0.94 &
  \cellcolor[HTML]{F4CCCC}49.2 &
  \cellcolor[HTML]{F4CCCC}486.0 &
  \cellcolor[HTML]{F4CCCC}0.93 &
  \cellcolor[HTML]{F4CCCC}51.1 &
  \cellcolor[HTML]{F4CCCC}480.2 &
  \cellcolor[HTML]{F4CCCC}0.92 &
  \cellcolor[HTML]{F4CCCC}56.8 &
  \cellcolor[HTML]{F4CCCC}1317.0 &
  \cellcolor[HTML]{F4CCCC}0.87 &
  \cellcolor[HTML]{F4CCCC}70.9 &
  \cellcolor[HTML]{C9DAF8}139.4 &
  \cellcolor[HTML]{C9DAF8}0.97 &
  \cellcolor[HTML]{C9DAF8}33.9 &
  \cellcolor[HTML]{C9DAF8}413.1 &
  \cellcolor[HTML]{C9DAF8}0.93 &
  \cellcolor[HTML]{C9DAF8}35.7 &
  \cellcolor[HTML]{C9DAF8}83.4 &
  \cellcolor[HTML]{C9DAF8}0.84 &
  \cellcolor[HTML]{C9DAF8}20.8 \\
\multirow{-5}{*}{HybridGNet} &
  \cellcolor[HTML]{D9EAD3}LHC (Full) &
  \cellcolor[HTML]{D9EAD3}110.7 &
  \cellcolor[HTML]{D9EAD3}0.97 &
  \cellcolor[HTML]{D9EAD3}26.2 &
  \cellcolor[HTML]{D9EAD3}148.7 &
  \cellcolor[HTML]{D9EAD3}0.96 &
  \cellcolor[HTML]{D9EAD3}33.2 &
  \cellcolor[HTML]{D9EAD3}172.0 &
  \cellcolor[HTML]{D9EAD3}0.96 &
  \cellcolor[HTML]{D9EAD3}37.1 &
  \cellcolor[HTML]{D9EAD3}235.0 &
  \cellcolor[HTML]{D9EAD3}0.94 &
  \cellcolor[HTML]{D9EAD3}30.5 &
  \cellcolor[HTML]{D9EAD3}122.2 &
  \cellcolor[HTML]{D9EAD3}0.97 &
  \cellcolor[HTML]{D9EAD3}31.1 &
  \cellcolor[HTML]{D9EAD3}390.8 &
  \cellcolor[HTML]{D9EAD3}0.94 &
  \cellcolor[HTML]{D9EAD3}34.2 &
  \cellcolor[HTML]{D9EAD3}101.1 &
  \cellcolor[HTML]{D9EAD3}0.82 &
  \cellcolor[HTML]{D9EAD3}22.9 \\ \hline
 &
  \cellcolor[HTML]{C9DAF8}L (Both) &
  \cellcolor[HTML]{C9DAF8}- &
  \cellcolor[HTML]{C9DAF8}0.98 &
  \cellcolor[HTML]{C9DAF8}53.3 &
  \cellcolor[HTML]{C9DAF8}- &
  \cellcolor[HTML]{C9DAF8}0.98 &
  \cellcolor[HTML]{C9DAF8}53.3 &
  \cellcolor[HTML]{C9DAF8}- &
  \cellcolor[HTML]{C9DAF8}0.96 &
  \cellcolor[HTML]{C9DAF8}85.2 &
  \cellcolor[HTML]{C9DAF8}- &
  \cellcolor[HTML]{C9DAF8}- &
  \cellcolor[HTML]{C9DAF8}- &
  \cellcolor[HTML]{C9DAF8}- &
  \cellcolor[HTML]{C9DAF8}0.95 &
  \cellcolor[HTML]{C9DAF8}99.9 &
  \cellcolor[HTML]{C9DAF8}- &
  \cellcolor[HTML]{C9DAF8}- &
  \cellcolor[HTML]{C9DAF8}- &
  \cellcolor[HTML]{C9DAF8}- &
  \cellcolor[HTML]{C9DAF8}- &
  \cellcolor[HTML]{C9DAF8}- \\
 &
  \cellcolor[HTML]{F4CCCC}LH (Strict) &
  \cellcolor[HTML]{F4CCCC}- &
  \cellcolor[HTML]{F4CCCC}0.96 &
  \cellcolor[HTML]{F4CCCC}73.4 &
  \cellcolor[HTML]{F4CCCC}- &
  \cellcolor[HTML]{F4CCCC}0.96 &
  \cellcolor[HTML]{F4CCCC}131.2 &
  \cellcolor[HTML]{C9DAF8}- &
  \cellcolor[HTML]{C9DAF8}0.96 &
  \cellcolor[HTML]{C9DAF8}89.4 &
  \cellcolor[HTML]{C9DAF8}- &
  \cellcolor[HTML]{C9DAF8}0.93 &
  \cellcolor[HTML]{C9DAF8}80.5 &
  \cellcolor[HTML]{C9DAF8}- &
  \cellcolor[HTML]{C9DAF8}0.95 &
  \cellcolor[HTML]{C9DAF8}101.1 &
  \cellcolor[HTML]{C9DAF8}- &
  \cellcolor[HTML]{C9DAF8}0.94 &
  \cellcolor[HTML]{C9DAF8}51.7 &
  \cellcolor[HTML]{C9DAF8}- &
  \cellcolor[HTML]{C9DAF8}- &
  \cellcolor[HTML]{C9DAF8}- \\
 &
  \cellcolor[HTML]{D9EAD3}LH (Full) &
  \cellcolor[HTML]{D9EAD3}- &
  \cellcolor[HTML]{D9EAD3}0.97 &
  \cellcolor[HTML]{D9EAD3}70.0 &
  \cellcolor[HTML]{D9EAD3}- &
  \cellcolor[HTML]{D9EAD3}0.96 &
  \cellcolor[HTML]{D9EAD3}74.0 &
  \cellcolor[HTML]{D9EAD3}- &
  \cellcolor[HTML]{D9EAD3}0.96 &
  \cellcolor[HTML]{D9EAD3}94.8 &
  \cellcolor[HTML]{D9EAD3}- &
  \cellcolor[HTML]{D9EAD3}0.89 &
  \cellcolor[HTML]{D9EAD3}75.2 &
  \cellcolor[HTML]{D9EAD3}- &
  \cellcolor[HTML]{D9EAD3}0.95 &
  \cellcolor[HTML]{D9EAD3}103.8 &
  \cellcolor[HTML]{D9EAD3}- &
  \cellcolor[HTML]{D9EAD3}0.94 &
  \cellcolor[HTML]{D9EAD3}51.8 &
  \cellcolor[HTML]{D9EAD3}- &
  \cellcolor[HTML]{D9EAD3}- &
  \cellcolor[HTML]{D9EAD3}- \\
 &
  \cellcolor[HTML]{F4CCCC}LHC (Strict) &
  \cellcolor[HTML]{F4CCCC}- &
  \cellcolor[HTML]{F4CCCC}0.90 &
  \cellcolor[HTML]{F4CCCC}182.4 &
  \cellcolor[HTML]{F4CCCC}- &
  \cellcolor[HTML]{F4CCCC}0.91 &
  \cellcolor[HTML]{F4CCCC}217.6 &
  \cellcolor[HTML]{F4CCCC}- &
  \cellcolor[HTML]{F4CCCC}0.89 &
  \cellcolor[HTML]{F4CCCC}227.1 &
  \cellcolor[HTML]{F4CCCC}- &
  \cellcolor[HTML]{F4CCCC}0.87 &
  \cellcolor[HTML]{F4CCCC}220.8 &
  \cellcolor[HTML]{C9DAF8}- &
  \cellcolor[HTML]{C9DAF8}0.97 &
  \cellcolor[HTML]{C9DAF8}67.5 &
  \cellcolor[HTML]{C9DAF8}- &
  \cellcolor[HTML]{C9DAF8}0.94 &
  \cellcolor[HTML]{C9DAF8}67.6 &
  \cellcolor[HTML]{C9DAF8}- &
  \cellcolor[HTML]{C9DAF8}0.93 &
  \cellcolor[HTML]{C9DAF8}44.8 \\
\multirow{-5}{*}{UNet} &
  \cellcolor[HTML]{D9EAD3}LHC (Full) &
  \cellcolor[HTML]{D9EAD3}- &
  \cellcolor[HTML]{D9EAD3}0.97 &
  \cellcolor[HTML]{D9EAD3}72.5 &
  \cellcolor[HTML]{D9EAD3}- &
  \cellcolor[HTML]{D9EAD3}0.97 &
  \cellcolor[HTML]{D9EAD3}72.7 &
  \cellcolor[HTML]{D9EAD3}- &
  \cellcolor[HTML]{D9EAD3}0.96 &
  \cellcolor[HTML]{D9EAD3}106.1 &
  \cellcolor[HTML]{D9EAD3}- &
  \cellcolor[HTML]{D9EAD3}0.91 &
  \cellcolor[HTML]{D9EAD3}76.7 &
  \cellcolor[HTML]{D9EAD3}- &
  \cellcolor[HTML]{D9EAD3}0.98 &
  \cellcolor[HTML]{D9EAD3}63.0 &
  \cellcolor[HTML]{D9EAD3}- &
  \cellcolor[HTML]{D9EAD3}0.94 &
  \cellcolor[HTML]{D9EAD3}54.5 &
  \cellcolor[HTML]{D9EAD3}- &
  \cellcolor[HTML]{D9EAD3}0.91 &
  \cellcolor[HTML]{D9EAD3}49.2 \\ \hline
 &
  \cellcolor[HTML]{C9DAF8}L (Both) &
  \cellcolor[HTML]{C9DAF8}- &
  \cellcolor[HTML]{C9DAF8}0.98 &
  \cellcolor[HTML]{C9DAF8}26.1 &
  \cellcolor[HTML]{C9DAF8}- &
  \cellcolor[HTML]{C9DAF8}0.97 &
  \cellcolor[HTML]{C9DAF8}32.3 &
  \cellcolor[HTML]{C9DAF8}- &
  \cellcolor[HTML]{C9DAF8}0.96 &
  \cellcolor[HTML]{C9DAF8}40.2 &
  \cellcolor[HTML]{C9DAF8}- &
  \cellcolor[HTML]{C9DAF8}- &
  \cellcolor[HTML]{C9DAF8}- &
  \cellcolor[HTML]{C9DAF8}- &
  \cellcolor[HTML]{C9DAF8}0.98 &
  \cellcolor[HTML]{C9DAF8}31.1 &
  \cellcolor[HTML]{C9DAF8}- &
  \cellcolor[HTML]{C9DAF8}- &
  \cellcolor[HTML]{C9DAF8}- &
  \cellcolor[HTML]{C9DAF8}- &
  \cellcolor[HTML]{C9DAF8}- &
  \cellcolor[HTML]{C9DAF8}- \\
 &
  \cellcolor[HTML]{F4CCCC}LH (Strict) &
  \cellcolor[HTML]{F4CCCC}- &
  \cellcolor[HTML]{F4CCCC}0.97 &
  \cellcolor[HTML]{F4CCCC}45.1 &
  \cellcolor[HTML]{F4CCCC}- &
  \cellcolor[HTML]{F4CCCC}0.96 &
  \cellcolor[HTML]{F4CCCC}53.3 &
  \cellcolor[HTML]{C9DAF8}- &
  \cellcolor[HTML]{C9DAF8}0.96 &
  \cellcolor[HTML]{C9DAF8}40.5 &
  \cellcolor[HTML]{C9DAF8}\textbf{-} &
  \cellcolor[HTML]{C9DAF8}0.95 &
  \cellcolor[HTML]{C9DAF8}32.9 &
  \cellcolor[HTML]{C9DAF8}- &
  \cellcolor[HTML]{C9DAF8}0.98 &
  \cellcolor[HTML]{C9DAF8}26.3 &
  \cellcolor[HTML]{C9DAF8}- &
  \cellcolor[HTML]{C9DAF8}0.95 &
  \cellcolor[HTML]{C9DAF8}29.0 &
  \cellcolor[HTML]{C9DAF8}- &
  \cellcolor[HTML]{C9DAF8}- &
  \cellcolor[HTML]{C9DAF8}- \\
 &
  \cellcolor[HTML]{D9EAD3}LH (Full) &
  \cellcolor[HTML]{D9EAD3}- &
  \cellcolor[HTML]{D9EAD3}0.98 &
  \cellcolor[HTML]{D9EAD3}34.0 &
  \cellcolor[HTML]{D9EAD3}- &
  \cellcolor[HTML]{D9EAD3}0.97 &
  \cellcolor[HTML]{D9EAD3}35.2 &
  \cellcolor[HTML]{D9EAD3}- &
  \cellcolor[HTML]{D9EAD3}0.96 &
  \cellcolor[HTML]{D9EAD3}37.4 &
  \cellcolor[HTML]{D9EAD3}- &
  \cellcolor[HTML]{D9EAD3}0.94 &
  \cellcolor[HTML]{D9EAD3}34.5 &
  \cellcolor[HTML]{D9EAD3}- &
  \cellcolor[HTML]{D9EAD3}0.98 &
  \cellcolor[HTML]{D9EAD3}28.6 &
  \cellcolor[HTML]{D9EAD3}- &
  \cellcolor[HTML]{D9EAD3}0.95 &
  \cellcolor[HTML]{D9EAD3}30.9 &
  \cellcolor[HTML]{D9EAD3}- &
  \cellcolor[HTML]{D9EAD3}- &
  \cellcolor[HTML]{D9EAD3}- \\
 &
  \cellcolor[HTML]{F4CCCC}LHC (Strict) &
  \cellcolor[HTML]{F4CCCC}- &
  \cellcolor[HTML]{F4CCCC}0.93 &
  \cellcolor[HTML]{F4CCCC}120.5 &
  \cellcolor[HTML]{F4CCCC}- &
  \cellcolor[HTML]{F4CCCC}0.92 &
  \cellcolor[HTML]{F4CCCC}121.0 &
  \cellcolor[HTML]{F4CCCC}- &
  \cellcolor[HTML]{F4CCCC}0.93 &
  \cellcolor[HTML]{F4CCCC}83.6 &
  \cellcolor[HTML]{F4CCCC}- &
  \cellcolor[HTML]{F4CCCC}0.89 &
  \cellcolor[HTML]{F4CCCC}80.4 &
  \cellcolor[HTML]{C9DAF8}- &
  \cellcolor[HTML]{C9DAF8}0.98 &
  \cellcolor[HTML]{C9DAF8}35.0 &
  \cellcolor[HTML]{C9DAF8}- &
  \cellcolor[HTML]{C9DAF8}0.95 &
  \cellcolor[HTML]{C9DAF8}29.5 &
  \cellcolor[HTML]{C9DAF8}- &
  \cellcolor[HTML]{C9DAF8}0.95 &
  \cellcolor[HTML]{C9DAF8}14.1 \\
\multirow{-5}{*}{nnUNet} &
  \cellcolor[HTML]{D9EAD3}LHC (Full) &
  \cellcolor[HTML]{D9EAD3}- &
  \cellcolor[HTML]{D9EAD3}0.98 &
  \cellcolor[HTML]{D9EAD3}25.3 &
  \cellcolor[HTML]{D9EAD3}- &
  \cellcolor[HTML]{D9EAD3}0.97 &
  \cellcolor[HTML]{D9EAD3}35.2 &
  \cellcolor[HTML]{D9EAD3}- &
  \cellcolor[HTML]{D9EAD3}0.96 &
  \cellcolor[HTML]{D9EAD3}39.3 &
  \cellcolor[HTML]{D9EAD3}- &
  \cellcolor[HTML]{D9EAD3}0.95 &
  \cellcolor[HTML]{D9EAD3}33.3 &
  \cellcolor[HTML]{D9EAD3}- &
  \cellcolor[HTML]{D9EAD3}0.98 &
  \cellcolor[HTML]{D9EAD3}35.6 &
  \cellcolor[HTML]{D9EAD3}- &
  \cellcolor[HTML]{D9EAD3}0.95 &
  \cellcolor[HTML]{D9EAD3}31.1 &
  \cellcolor[HTML]{D9EAD3}- &
  \cellcolor[HTML]{D9EAD3}0.93 &
  \cellcolor[HTML]{D9EAD3}38.1 \\ \hline
 &
  \cellcolor[HTML]{C9DAF8}L (Both) &
  \cellcolor[HTML]{C9DAF8}- &
  \cellcolor[HTML]{C9DAF8}0.97 &
  \cellcolor[HTML]{C9DAF8}46.9 &
  \cellcolor[HTML]{C9DAF8}- &
  \cellcolor[HTML]{C9DAF8}0.97 &
  \cellcolor[HTML]{C9DAF8}78.7 &
  \cellcolor[HTML]{C9DAF8}- &
  \cellcolor[HTML]{C9DAF8}0.96 &
  \cellcolor[HTML]{C9DAF8}70.0 &
  \cellcolor[HTML]{C9DAF8}- &
  \cellcolor[HTML]{C9DAF8}- &
  \cellcolor[HTML]{C9DAF8}- &
  \cellcolor[HTML]{C9DAF8}- &
  \cellcolor[HTML]{C9DAF8}0.95 &
  \cellcolor[HTML]{C9DAF8}106.1 &
  \cellcolor[HTML]{C9DAF8}- &
  \cellcolor[HTML]{C9DAF8}- &
  \cellcolor[HTML]{C9DAF8}- &
  \cellcolor[HTML]{C9DAF8}- &
  \cellcolor[HTML]{C9DAF8}- &
  \cellcolor[HTML]{C9DAF8}- \\
 &
  \cellcolor[HTML]{F4CCCC}LH (Strict) &
  \cellcolor[HTML]{F4CCCC}- &
  \cellcolor[HTML]{F4CCCC}0.96 &
  \cellcolor[HTML]{F4CCCC}74.8 &
  \cellcolor[HTML]{F4CCCC}- &
  \cellcolor[HTML]{F4CCCC}0.96 &
  \cellcolor[HTML]{F4CCCC}131.6 &
  \cellcolor[HTML]{C9DAF8}- &
  \cellcolor[HTML]{C9DAF8}0.96 &
  \cellcolor[HTML]{C9DAF8}74.9 &
  \cellcolor[HTML]{C9DAF8}- &
  \cellcolor[HTML]{C9DAF8}0.94 &
  \cellcolor[HTML]{C9DAF8}80.2 &
  \cellcolor[HTML]{C9DAF8}- &
  \cellcolor[HTML]{C9DAF8}0.95 &
  \cellcolor[HTML]{C9DAF8}105.6 &
  \cellcolor[HTML]{C9DAF8}- &
  \cellcolor[HTML]{C9DAF8}0.94 &
  \cellcolor[HTML]{C9DAF8}58.4 &
  \cellcolor[HTML]{C9DAF8}- &
  \cellcolor[HTML]{C9DAF8}- &
  \cellcolor[HTML]{C9DAF8}- \\
 &
  \cellcolor[HTML]{D9EAD3}LH (Full) &
  \cellcolor[HTML]{D9EAD3}- &
  \cellcolor[HTML]{D9EAD3}0.98 &
  \cellcolor[HTML]{D9EAD3}60.5 &
  \cellcolor[HTML]{D9EAD3}- &
  \cellcolor[HTML]{D9EAD3}0.97 &
  \cellcolor[HTML]{D9EAD3}57.8 &
  \cellcolor[HTML]{D9EAD3}- &
  \cellcolor[HTML]{D9EAD3}0.96 &
  \cellcolor[HTML]{D9EAD3}87.9 &
  \cellcolor[HTML]{D9EAD3}- &
  \cellcolor[HTML]{D9EAD3}0.93 &
  \cellcolor[HTML]{D9EAD3}125.6 &
  \cellcolor[HTML]{D9EAD3}- &
  \cellcolor[HTML]{D9EAD3}0.95 &
  \cellcolor[HTML]{D9EAD3}100.2 &
  \cellcolor[HTML]{D9EAD3}- &
  \cellcolor[HTML]{D9EAD3}0.94 &
  \cellcolor[HTML]{D9EAD3}59.4 &
  \cellcolor[HTML]{D9EAD3}- &
  \cellcolor[HTML]{D9EAD3}- &
  \cellcolor[HTML]{D9EAD3}- \\
 &
  \cellcolor[HTML]{F4CCCC}LHC (Strict) &
  \cellcolor[HTML]{F4CCCC}- &
  \cellcolor[HTML]{F4CCCC}0.91 &
  \cellcolor[HTML]{F4CCCC}168.5 &
  \cellcolor[HTML]{F4CCCC}- &
  \cellcolor[HTML]{F4CCCC}0.91 &
  \cellcolor[HTML]{F4CCCC}204.8 &
  \cellcolor[HTML]{F4CCCC}- &
  \cellcolor[HTML]{F4CCCC}0.90 &
  \cellcolor[HTML]{F4CCCC}223.6 &
  \cellcolor[HTML]{F4CCCC}- &
  \cellcolor[HTML]{F4CCCC}0.87 &
  \cellcolor[HTML]{F4CCCC}199.0 &
  \cellcolor[HTML]{C9DAF8}- &
  \cellcolor[HTML]{C9DAF8}0.98 &
  \cellcolor[HTML]{C9DAF8}78.1 &
  \cellcolor[HTML]{C9DAF8}- &
  \cellcolor[HTML]{C9DAF8}0.94 &
  \cellcolor[HTML]{C9DAF8}82.2 &
  \cellcolor[HTML]{C9DAF8}- &
  \cellcolor[HTML]{C9DAF8}0.94 &
  \cellcolor[HTML]{C9DAF8}28.3 \\
\multirow{-5}{*}{UNet HT} &
  \cellcolor[HTML]{D9EAD3}LHC (Full) &
  \cellcolor[HTML]{D9EAD3}- &
  \cellcolor[HTML]{D9EAD3}0.97 &
  \cellcolor[HTML]{D9EAD3}72.6 &
  \cellcolor[HTML]{D9EAD3}- &
  \cellcolor[HTML]{D9EAD3}0.97 &
  \cellcolor[HTML]{D9EAD3}77.1 &
  \cellcolor[HTML]{D9EAD3}- &
  \cellcolor[HTML]{D9EAD3}0.96 &
  \cellcolor[HTML]{D9EAD3}66.1 &
  \cellcolor[HTML]{D9EAD3}- &
  \cellcolor[HTML]{D9EAD3}0.93 &
  \cellcolor[HTML]{D9EAD3}87.8 &
  \cellcolor[HTML]{D9EAD3}- &
  \cellcolor[HTML]{D9EAD3}0.98 &
  \cellcolor[HTML]{D9EAD3}55.2 &
  \cellcolor[HTML]{D9EAD3}- &
  \cellcolor[HTML]{D9EAD3}0.94 &
  \cellcolor[HTML]{D9EAD3}47.6 &
  \cellcolor[HTML]{D9EAD3}- &
  \cellcolor[HTML]{D9EAD3}0.94 &
  \cellcolor[HTML]{D9EAD3}43.6 \\ \hline
\end{tabular}}

\caption{\textbf{Quantitative results for both landmark and pixel-based baselines:} Results show an increase in performance when combining heterogeneous labels (Full) compared to those cases where only images with all the required annotations (Strict) are available. \textbf{Blue:} images from the target dataset are present at training time. \textbf{Red:} images from the target dataset are not present during training. \textbf{Green:} heterogeneous setting, all datasets are present during training. 
}

\label{table:table_nnunet}

\end{table*}

%% file: table2.tex
\begin{table}[]
\resizebox{0.5\textwidth}{!}{
\begin{tabular}{cc|cccc|cccc}
\hline
 &
   &
  \multicolumn{4}{c|}{\textbf{JSRT}} &
  \multicolumn{4}{c}{\textbf{Padchest}} \\
 &
   &
  \multicolumn{2}{c}{\textbf{Lungs}} &
  \multicolumn{2}{c|}{\textbf{Heart}} &
  \multicolumn{2}{c}{\textbf{Lungs}} &
  \multicolumn{2}{c}{\textbf{Heart}} \\
\multirow{-3}{*}{\textbf{Model}} &
  \multirow{-3}{*}{\textbf{Trained in}} &
  \textbf{Dice} &
  \textbf{HD} &
  \textbf{Dice} &
  \textbf{HD} &
  \textbf{Dice} &
  \textbf{HD} &
  \textbf{Dice} &
  \textbf{HD} \\ \hline
 &
  Exp 1 &
  \cellcolor[HTML]{F4CCCC}0.931 &
  \cellcolor[HTML]{F4CCCC}47.2 &
  \cellcolor[HTML]{C9DAF8}0.941 &
  \cellcolor[HTML]{C9DAF8}31.2 &
  \cellcolor[HTML]{C9DAF8}0.952 &
  \cellcolor[HTML]{C9DAF8}37.8 &
  \cellcolor[HTML]{C9DAF8}0.942 &
  \cellcolor[HTML]{C9DAF8}31.4 \\
 &
  Exp 2 &
  \cellcolor[HTML]{C9DAF8}0.970 &
  \cellcolor[HTML]{C9DAF8}34.0 &
  \cellcolor[HTML]{F4CCCC}0.910 &
  \cellcolor[HTML]{F4CCCC}58.2 &
  \cellcolor[HTML]{C9DAF8}0.953 &
  \cellcolor[HTML]{C9DAF8}40.5 &
  \cellcolor[HTML]{C9DAF8}0.935 &
  \cellcolor[HTML]{C9DAF8}33.7 \\
 &
  Exp 3 &
  \cellcolor[HTML]{C9DAF8}0.970 &
  \cellcolor[HTML]{C9DAF8}34.3 &
  \cellcolor[HTML]{C9DAF8}0.941 &
  \cellcolor[HTML]{C9DAF8}32.3 &
  \cellcolor[HTML]{F4CCCC}0.905 &
  \cellcolor[HTML]{F4CCCC}56.3 &
  \cellcolor[HTML]{C9DAF8}0.946 &
  \cellcolor[HTML]{C9DAF8}31.4 \\
\multirow{-4}{*}{HybridGNet} &
  Exp 4 &
  \cellcolor[HTML]{C9DAF8}0.975 &
  \cellcolor[HTML]{C9DAF8}32.6 &
  \cellcolor[HTML]{C9DAF8}0.939 &
  \cellcolor[HTML]{C9DAF8}31.9 &
  \cellcolor[HTML]{C9DAF8}0.957 &
  \cellcolor[HTML]{C9DAF8}38.1 &
  \cellcolor[HTML]{F4CCCC}0.899 &
  \cellcolor[HTML]{F4CCCC}69.7 \\ \hline
 &
  Exp 1 &
  \cellcolor[HTML]{F4CCCC}0.968 &
  \cellcolor[HTML]{F4CCCC}90.3 &
  \cellcolor[HTML]{C9DAF8}0.937 &
  \cellcolor[HTML]{C9DAF8}80.0 &
  \cellcolor[HTML]{C9DAF8}0.960 &
  \cellcolor[HTML]{C9DAF8}96.9 &
  \cellcolor[HTML]{C9DAF8}0.928 &
  \cellcolor[HTML]{C9DAF8}141.9 \\
 &
  Exp 2 &
  \cellcolor[HTML]{C9DAF8}0.980 &
  \cellcolor[HTML]{C9DAF8}57.3 &
  \cellcolor[HTML]{F4CCCC}0.860 &
  \cellcolor[HTML]{F4CCCC}340.6 &
  \cellcolor[HTML]{C9DAF8}0.960 &
  \cellcolor[HTML]{C9DAF8}69.6 &
  \cellcolor[HTML]{C9DAF8}0.923 &
  \cellcolor[HTML]{C9DAF8}116.6 \\
 &
  Exp 3 &
  \cellcolor[HTML]{C9DAF8}0.979 &
  \cellcolor[HTML]{C9DAF8}44.0 &
  \cellcolor[HTML]{C9DAF8}0.941 &
  \cellcolor[HTML]{C9DAF8}46.7 &
  \cellcolor[HTML]{F4CCCC}0.932 &
  \cellcolor[HTML]{F4CCCC}202.0 &
  \cellcolor[HTML]{C9DAF8}0.931 &
  \cellcolor[HTML]{C9DAF8}102.7 \\
\multirow{-4}{*}{UNet HT} &
  Exp 4 &
  \cellcolor[HTML]{C9DAF8}0.979 &
  \cellcolor[HTML]{C9DAF8}53.4 &
  \cellcolor[HTML]{C9DAF8}0.941 &
  \cellcolor[HTML]{C9DAF8}66.7 &
  \cellcolor[HTML]{C9DAF8}0.960 &
  \cellcolor[HTML]{C9DAF8}63.3 &
  \cellcolor[HTML]{F4CCCC}0.872 &
  \cellcolor[HTML]{F4CCCC}182.5 \\ \hline
 &
  Exp 1 &
  \cellcolor[HTML]{F4CCCC}0.034 &
  \cellcolor[HTML]{F4CCCC}- &
  \cellcolor[HTML]{C9DAF8}0.937 &
  \cellcolor[HTML]{C9DAF8}56.5 &
  \cellcolor[HTML]{C9DAF8}0.956 &
  \cellcolor[HTML]{C9DAF8}59.3 &
  \cellcolor[HTML]{C9DAF8}0.935 &
  \cellcolor[HTML]{C9DAF8}82.3 \\
 &
  Exp 2 &
  \cellcolor[HTML]{C9DAF8}0.979 &
  \cellcolor[HTML]{C9DAF8}65.8 &
  \cellcolor[HTML]{F4CCCC}0.027 &
  \cellcolor[HTML]{F4CCCC}- &
  \cellcolor[HTML]{C9DAF8}0.959 &
  \cellcolor[HTML]{C9DAF8}98.5 &
  \cellcolor[HTML]{C9DAF8}0.936 &
  \cellcolor[HTML]{C9DAF8}101.5 \\
 &
  Exp 3 &
  \cellcolor[HTML]{C9DAF8}0.977 &
  \cellcolor[HTML]{C9DAF8}56.1 &
  \cellcolor[HTML]{C9DAF8}0.938 &
  \cellcolor[HTML]{C9DAF8}71.4 &
  \cellcolor[HTML]{F4CCCC}0.035 &
  \cellcolor[HTML]{F4CCCC}-- &
  \cellcolor[HTML]{C9DAF8}0.921 &
  \cellcolor[HTML]{C9DAF8}153.7 \\
\multirow{-4}{*}{UNet} &
  Exp 4 &
  \cellcolor[HTML]{C9DAF8}0.980 &
  \cellcolor[HTML]{C9DAF8}49.0 &
  \cellcolor[HTML]{C9DAF8}0.944 &
  \cellcolor[HTML]{C9DAF8}68.4 &
  \cellcolor[HTML]{C9DAF8}0.961 &
  \cellcolor[HTML]{C9DAF8}95.0 &
  \cellcolor[HTML]{F4CCCC}0.039 &
  \cellcolor[HTML]{F4CCCC}- \\ \hline
 &
  Exp 1 &
  \cellcolor[HTML]{F4CCCC}0.000 &
  \cellcolor[HTML]{F4CCCC}- &
  \cellcolor[HTML]{C9DAF8}0.952 &
  \cellcolor[HTML]{C9DAF8}30.2 &
  \cellcolor[HTML]{C9DAF8}0.963 &
  \cellcolor[HTML]{C9DAF8}43.4 &
  \cellcolor[HTML]{C9DAF8}0.947 &
  \cellcolor[HTML]{C9DAF8}31.6 \\
 &
  Exp 2 &
  \cellcolor[HTML]{C9DAF8}0.980 &
  \cellcolor[HTML]{C9DAF8}33.4 &
  \cellcolor[HTML]{F4CCCC}0.000 &
  \cellcolor[HTML]{F4CCCC}- &
  \cellcolor[HTML]{C9DAF8}0.965 &
  \cellcolor[HTML]{C9DAF8}38.8 &
  \cellcolor[HTML]{C9DAF8}0.945 &
  \cellcolor[HTML]{C9DAF8}37.8 \\
 &
  Exp 3 &
  \cellcolor[HTML]{C9DAF8}0.979 &
  \cellcolor[HTML]{C9DAF8}47.4 &
  \cellcolor[HTML]{C9DAF8}0.948 &
  \cellcolor[HTML]{C9DAF8}30.1 &
  \cellcolor[HTML]{F4CCCC}0.000 &
  \cellcolor[HTML]{F4CCCC}- &
  \cellcolor[HTML]{C9DAF8}0.945 &
  \cellcolor[HTML]{C9DAF8}31.7 \\
\multirow{-4}{*}{nnUNet} &
  Exp 4 &
  \cellcolor[HTML]{C9DAF8}0.981 &
  \cellcolor[HTML]{C9DAF8}32.1 &
  \cellcolor[HTML]{C9DAF8}0.951 &
  \cellcolor[HTML]{C9DAF8}29.6 &
  \cellcolor[HTML]{C9DAF8}0.964 &
  \cellcolor[HTML]{C9DAF8}38.2 &
  \cellcolor[HTML]{F4CCCC}0.000 &
  \cellcolor[HTML]{F4CCCC}- \\ \hline
\end{tabular}}

\caption{\textbf{Artificially removed labels experiment.} 
\textbf{In red:} label was not present during training. \textbf{In blue:} label was present during training.
}

\label{table:table_experiment}

\end{table}